\documentclass[a4paper]{JHEP3}

\usepackage[latin1]{inputenc}
\usepackage{bbm}
\usepackage{bm}
\usepackage{graphicx}
\usepackage{epsfig}
\usepackage{subfigure}
\usepackage{latexsym}
\usepackage{amsmath}
\usepackage{amsfonts}
\usepackage{amssymb}

\newcommand{\tr}{\textrm{Tr}}

\newcommand{\be}{\begin{eqnarray}} % only untightened
\newcommand{\ee}{\end{eqnarray}}

\newcommand{\bmp}{\noindent\begin{minipage}{16cm}}
\newcommand{\emp}{\end{minipage}\vskip 7mm} % 7mm untightened
\def\lsim{\mathrel{\raise.3ex\hbox{$<$\kern-.75em\lower1ex\hbox{$\sim$}}}}
\def\gsim{\mathrel{\raise.3ex\hbox{$>$\kern-.75em\lower1ex\hbox{$\sim$}}}}

\newcommand{\intron}[1]{}%{#1}
\title{Unnatural Origin of Fermion Masses for Technicolor}
\preprint{\it CP$^3$-Origins: 2009-18}
\author{Matti Antola\footnote{matti.antola@helsinki.fi}\\
Department of Physics and Helsinki Institute of Physics, 
P.O.Box 64, FI-000140, University of Helsinki, Finland}
\author{Matti Heikinheimo\footnote{matti.p.s.heikinheimo@jyu.fi}\\
Department of Physics, 
P.O.Box 35, FI-000140, University of Jyv\"askyl\"a and \\Helsinki Institute of Physics, 
P.O.Box 64, FI-000140, University of Helsinki, Finland}
\author{Francesco Sannino\footnote{sannino@cp3.sdu.dk}\\
{ CP}$^{ \bf 3}${-Origins}, 
Campusvej 55, DK-5230 Odense M, Denmark}
\author{Kimmo Tuominen \footnote{kimmo.tuominen@jyu.fi}\,\,\footnote{On leave of absence from Department of physics, University of Jyv\"askyl\"a} \\
{ CP}$^{ \bf 3}${-Origins}, 
Campusvej 55, DK-5230 Odense M, Denmark\footnote{{ C}entre of Excellence for { P}article { P}hysics { P}henomenology devoted to the understanding of the {Origins} of Mass in the universe} and\\
Helsinki Institute of Physics, 
P.O.Box 64, FI-000140, University of Helsinki, Finland}
\abstract
{We explore the scenario in which the breaking of the electroweak symmetry is due to the simultaneous presence and interplay of a dynamical sector and an unnatural elementary Higgs. We introduce a low energy effective  Lagrangian and constrain the various couplings via direct search limits and electroweak and flavor precision tests. We find that the model we study is a viable model of dynamical breaking of the electroweak symmetry.}

\maketitle

\begin{document}

\section{Introduction}
\label{intro}

Recently models of dynamical electroweak symmetry breaking \cite{TC,Hill:2002ap, Sannino:2008ha}  are receiving much attention thanks to progress made towards the understanding of the phase diagram of strongly coupled theories as function of the number of flavors,  colors and matter representation  \cite{Sannino:2004qp,Dietrich:2005jn,Dietrich:2006cm,Ryttov:2007cx,Sannino:2009aw}.  The breakthrough with respect to earlier models has been the realization that one can achieve near conformal dynamics with a very small number of flavors using matter in higher dimensional representations \cite{Sannino:2004qp,Dietrich:2005jn,Dietrich:2006cm,Ryttov:2007cx,Sannino:2009aw} or combinations of different representations \cite{Ryttov:2008xe,Ryttov:2009yw}. An exciting possibility is that different technicolor gauge theories are related by a gauge-duality transformation suggested recently in \cite{Sannino:2009qc,Sannino:2009me}. These new findings are compatible with the conjectured form of the beta function proposed in \cite{Ryttov:2007cx,Sannino:2009aw}; for recent extensions, see \cite{Antipin:2009wr}. The models with higher representation matter fields have been shown to be compatible with the current precision data \cite{Dietrich:2005jn,Foadi:2007ue,Foadi:2007se}. Their phenomenological implications for LHC experiments \cite{Belyaev:2008yj,Zerwekh:2005wh,Foadi:2008ci,Antipin:2009ks,Frandsen:2009fs,Antipin:2009ch} as well as for cosmology \cite{Foadi:2008qv,Kainulainen:2006wq,Jarvinen:2009pk} are being investigated. Recently the non-perturbative properties of these models have been  investigated on the lattice \cite{Catterall:2007yx,DelDebbio:2008wb,Catterall:2008qk,Hietanen:2008mr,Hietanen:2009az,Shamir:2008pb,DelDebbio:2009fd,Fodor:2009ar,Bursa:2009tj,Sinclair:2009ec}  

It is well known that to endow the standard model (SM) fermions with masses one has to extend the technicolor sector.  A popular approach has been to insist that there are no fundamental scalars in nature and hence new strong dynamics featuring only fermionic matter would lead to the generation of ordinary fermion masses 
\cite{Dimopoulos:1979es,Eichten:1979ah}. Despite the many efforts and the several models introduced in the literature it is fair to say  that none of them stands out as a particularly compelling setup. Reference \cite{Hill:2002ap} reviews the array of models existing up till 2003. It would be highly desirable to encode, at least partially, the effects of this sector within a well defined and calculable framework. One extremely economical way to achieve this goal has been to assume that the Yukawa sector of the SM represents the yet unknown new dynamics responsible to give mass to the SM fermions. This is the route followed in \cite{Foadi:2007ue,Fukano:2009zm}. This approach has the advantage of being very simple and predictive although it lacks the interplay between the technicolor and the sector giving masses to the SM fermions. In this work we investigate this type of interplay.

To achieve this goal we adopt a straightforward and instructive model according to which we have both a composite sector and  a fundamental scalar field  (SM-like Higgs) intertwined at the electroweak scale. This idea was pioneered  in a series of papers by Simmons  \cite{Simmons:1988fu},  Kagan and Samuel \cite{Kagan:1991gh} and Carone and Georgi  \cite{Carone:1992rh,Carone:1994mx}. More recently this type of model has been investigated also in \cite{Hemmige:2001vq,Carone:2006wj,Zerwekh:2009yu}. Interesting related work can be also found in \cite{Chivukula:1990bc,Chivukula:2009ck}. 

In the earlier work it was noted that these models permit to write renormalizable Yukawa interactions with ordinary fermions replacing, de facto, the extended technicolor dynamics. In comparison to the earlier works we have: 
\begin{itemize}
\item Included all dimension four operators with at most one mixing between the two scalar sectors. 
\item Provided an extensive scan of the parameters of the model. 
\item Updated the comparison with measurements. 
\item We linked the dynamical sector with models of  (Ultra) Minimal Walking technicolor \cite{Sannino:2004qp,Dietrich:2005jn,Dietrich:2006cm,Ryttov:2008xe}.  
\end{itemize}
An important point of this model is that no need for large anomalous dimensions of the technicolor condensate is needed. However, the near conformal dynamics is still relevant to reduce the contributions to the electroweak precision data. This makes the minimal walking technicolor models interesting for this kind of model building. 

The starting point is a low energy effective theory containing a composite Higgs ($M$) and an elementary one ($H$). Both fields transform according to the $SU(2)\times SU(2)$ global symmetry group and its maximal diagonal subgroup corresponds to the SM custodial symmetry.  Identifying $M$ with the composite degree of freedom, and in the absence of explicit extended technicolor interactions, we forbid its Yukawa couplings with SM fermions.  On the other hand the elementary $H$ field has a direct coupling to the SM fermions.

{}We will assume that the underlying gauge dynamics for the composite sector  is the Next to Minimal Walking Technicolor (NMWT) model  \cite{Dietrich:2005jn}.  Thanks to this choice we know that, in the absence of $H$,  the technicolor sector does not deviate substantially from the electroweak precision data. While other possibilities, like the minimal or ultraminimal walking technicolor \cite{Ryttov:2008xe} could be considered, NMWT has the simplest global symmetry structure, so we chose it as the starting point for building models of this type. 

Seeking a more natural justification for such models, one can imagine, for example, that supersymmetry is realized in Nature at some scale higher than the electroweak one along the lines suggested in \cite{Dine:1981za}. Interestingly, one can also investigate  possible grand unified models featuring technicolor dynamics at the electroweak scale \cite{Gudnason:2006mk}.

We will introduce the model in the next section and then investigate precision and flavor constraints in the rest of the paper. We find that this model is phenomenologically viable.  

\section{The Model}
\label{model}

The content of the model is described by the following Lagrangian:
\begin{equation}
\mathcal{L}_{UTC}=\mathcal{L}_{SM}\Bigr|_{\rm{Higgs}=0}+\mathcal{L}_{TC}+\mathcal{L}_{\rm{Higgs}}+\mathcal{L}_{\rm{Yukawa}}\;.
\label{master_lagrangian}
\end{equation}
We have fully separated the Higgs sector from the SM to emphasize that we are also coupling the technicolor sector to the fundamental scalar doublet through Yukawa couplings. Since the global chiral symmetry of the NMWT is SU(2)$\times$SU(2), the low energy effective Lagrangian treatment can be applied also to other models of strong dynamics with the same global symmetry, making our analysis more generic. The particle content of the NMWT model consists of $N_{f}=2$ (number of techniflavors) in the two-index symmetric representation of the technicolor gauge group, SU$(3)$. The two flavors are arranged into one doublet of SU$(2)_{\text{L}}$, and taking the technicolor degree of freedom into account implies that we are adding six doublets of SU$(2)_{\text{L}}$ and there is no Witten anomaly. The gauge structure of the whole model is SU$(3)_{\text{TC}}\times$SU$(3)_{\text{C}}\times$SU$(2)_{\text{L}}\times$U$(1)_{\text{Y}}$
and the anomaly free hypercharge assignments for the Techniquarks are
\be
Y(U_L)=Y(D_L)=0,~~Y(U_R)=\frac{1}{2},~~Y(D_R)=-\frac{1}{2}.
\ee

Next, let us explicitly construct the linearly realized effective Lagrangian of this model, which will be used throughout the rest of this article. The effective variable of the TC sector is taken to be the matrix
\begin{equation}
M=\frac{1}{\sqrt{2}}\left(sI_{2\times 2}+2i{\pi}_{M}\right)\propto Q_{L}\bar{Q}_{R},\;\;\;\;\left\langle s\right\rangle \equiv f
\end{equation}
and for notational simplicity, the Higgs field is also written in an analogous form:
\begin{equation}
H=\frac{1}{\sqrt{2}}\left(hI_{2x2}+2i{\pi}_{H}\right),\;\;\;\;\left\langle h\right\rangle \equiv v\;.
\end{equation}
We denote ${\pi}\equiv \pi^{i}\tau_{i}$, where $\tau_{i}$ are the SU(2) generators, i.e. $\tau_i=\sigma_i/2$ in terms of the Pauli matrices $\sigma_i$.

If we let the Higgs field transform under the gauge group SU$(2)_{\text{L}}\times$U$(1)_{\text{Y}}$ as $H\rightarrow g_{L}Hg_{Y}^{\dagger}$, where the U$(1)$ is generated by $\tau_3$, the explicit form of the Higgs Lagrangian in (\ref{master_lagrangian}) can be deduced:
\be
\mathcal{L}_{\rm{Higgs}}=\frac{1}{2}Tr\left[DH^{\dagger}DH\right]-V_{H},\; V_{H}=\frac{1}{2}m^{2}_HTr\left[H^{\dagger}H\right]+\frac{\lambda_{H}}{4!}Tr^{2}\left[H^{\dagger}H\right]
\label{Higgs_lagrangian}
\ee
In order to construct an effective Lagrangian with the fields $M$ and $H$ and with the correct symmetries and symmetry breaking properties, we must now determine the global symmetry structure of the theory. It is clear that with gauge fields turned off, the Higgs sector is invariant under a global SU$(2)_{\text{L}}\times$SU$(2)_{\text{R}}$ under which $H\rightarrow L_{H}HR_{H}^{\dagger}$. In addition to the global symmetry of the Higgs sector, the NMWT sector has a separate U$(1)_{V}\times$SU$(2)_{\text{L}}\times$SU$(2)_{\text{R}}$ chiral symmetry, under which $Q_{L}\rightarrow \exp(i\alpha)L_{TC}Q_{L}$ and $Q_{R}\rightarrow \exp(i\alpha)R_{TC}Q_{R}$. The subscripts TC and H refer to the fact that these symmetries originate from different SU(2)$\times$SU(2) symmetries associated with the TC and fundamental Higgs sectors. Thus, in the absence of gauge fields, the largest global symmetry of the kinetic terms is U$(1)_{V}\times($SU$(2)_{\text{L}}\times$SU$(2)_{\text{R}})^{2}$.

The largest possible symmetry of the kinetic terms, discussed above, is broken by two sources: gauge interactions and Yukawa interactions. We include the effects of the Yukawa interactions in the effective Lagrangian via the spurion technique. The Yukawa terms, denoted by ${\mathcal{L}}_{\rm{Yukawa}}$ in (\ref{master_lagrangian}), are of the form
\begin{eqnarray}
\mathcal{L}_{\rm{Yukawa}} & = & -\sum_{i=q,l,Q}\bar{\Psi}_{L,i}HY^{i}\Psi_{R,i}\;\;,\;\;Y^{i}=y^{i}I_{SU(2)}+\delta y^{i}\sigma_{3}.
\end{eqnarray}
where the sum is over SM quarks, leptons, and also techniquarks. Of current interest is the Yukawa term of the techniquarks:
\begin{eqnarray}
 -\bar{Q}_{L}HY_{Q}Q_{R}\label{Yukawa1},
\end{eqnarray}
The coupling (\ref{Yukawa1}) breaks $($SU$(2)_{\text{L}})^{2}\rightarrow$SU$(2)_{\text{L}}$ and $($SU$(2)_{\text{R}})^{2}\rightarrow$U$(1)_{\text{R}}$, where U$(1)_{\text{R}}$ is the diagonal subgoup generated by $\tau_3$. Thus, to preserve the TC chiral symmetry, the spurion field $HY_Q$ should transform as $HY_Q\rightarrow L_{\text{TC}}HY_QR_{\text{TC}}^{\dagger}$.  When the two origins of  global symmetry breaking are considered simultaneously, i.e. the Yukawa interactions and the electroweak gauging, the left over global symmetry is just U$(1)_V$. 

Because we have in mind a well defined underlying technicolor gauge theory we see that the global symmetries are different than the ones arising in a generic two Higgs doublet model (2HDM). In the latter case there is no notion of $U(1)_V$ baryon number given that there are no technibaryons in this theory. Also, in our case FCNC interactions are automatically reduced because of the symmetry breaking pattern, so we do not need to impose an additional $Z_2$ symmetry on one of the fields as is often done in the case of 2HDM phenomenology. As a matter of fact the sum $V_M+V_H$ does not admit such a symmetry.

We will also apply Georgi's generalized naive dimensional analysis \cite{Georgi:1992dw} to estimate the coefficients of operators. According to this counting  the coefficients depend on only two dimensionful quantities: $f$, which is the Goldstone boson decay constant, and $\Lambda$, the mass of some low lying non-Goldstone state. By interpreting $1/f$ as a universal measure of the amplitude for producing a strongly interacting bound state, the rules are:
\begin{itemize}
\item each strongly interacting field appears in the Lagrangian accompanied by a factor of $1/f$ 
\item multiply each Lagrangian term by a factor of $f^{2}$
\item fix the overall dimension of each term by multiplying by the appropriate factors of $\Lambda$
\end{itemize}
The naive counting relies on assuming the ratio $\alpha = \Lambda/f$ to be greater than one (say of the order of $4\pi$).

We include all possible terms up to and including dimension four operators consistent with symmetries, but to limit the amount of terms, we assume that $Y_Q$ is not too large and omit terms of order $\mathcal{O}(Y_Q^{2})$ and higher. We arrive at the following effective low-energy Lagrangian for the Technicolor sector and its coupling with the fundamental Higgs:
\be
\mathcal{L}_{TC}-\bar{Q}_{L}HY_QQ_{R} & \rightarrow &
\frac{1}{2}\tr\left[DM^{\dagger}DM\right]+\frac{1}{2}(c_{3}/\alpha)\tr\left[DM^{\dagger}DHY_Q\right]-V_{M}\nonumber \\
V_{M} &=& \frac{1}{2}m^{2}_M\tr\left[M^{\dagger}M\right]+\frac{\lambda_{M}}{4!}\tr^{2}\left[M^{\dagger}M\right]\nonumber \\
 & & -\frac{1}{2}(\alpha c_{1})f^{2}\tr\left[M^{\dagger}HY_Q\right]-\frac{1}{24}(\alpha c_{2})\tr\left[M^{\dagger}M\right]\tr\left[M^{\dagger}HY_Q\right]\nonumber \\
 & & -\frac{1}{24}(c_{4}/\alpha)\tr\left[H^{\dagger}H\right]\tr\left[M^{\dagger}HY_Q\right]+\text{h.c.}
\label{TC_lagrangian}
\ee
The dimensionless coefficients $c_{1}$ ... $c_{4}$ are order one quantities and  taken to be real to preserve the CP symmetry. 

Consider next the kinetic part of both the fundamental scalar (Higgs) from (\ref{Higgs_lagrangian}) and the composites (technicolor) from (\ref{TC_lagrangian})
\begin{equation}
\mathcal{L}_{KE}=\frac{1}{2}\tr\left[DM^{\dagger}DM\right]+\frac{1}{2}\tr\left[DH^{\dagger}DH\right]+\frac{1}{2}(c_{3}/\alpha)\tr\left[\left(DM^{\dagger}DH+DH^{\dagger}DM\right)Y_Q\right]
\label{eq:KEmix}
\end{equation}
As can be seen by a direct computation, the $\sigma_{3}$-part of $Y_Q=y_{Q}+\delta y_{Q}\sigma_{3}$ does not contribute, and therefore we can diagonalize the kinetic terms by inserting (note that the subindex $\pm$ is a label and does not indicate electric charge)\begin{equation}
\left(\begin{array}{c}
M\\
H\end{array}\right)=\frac{1}{\sqrt{2}}\left(\begin{array}{cc}
\frac{1}{\sqrt{1-(c_{3}/\alpha)y_{Q}}} & \frac{1}{\sqrt{1+(c_{3}/\alpha)y_{Q}}}\\
-\frac{1}{\sqrt{1-(c_{3}/\alpha)y_{Q}}} & \frac{1}{\sqrt{1+(c_{3}/\alpha)y_{Q}}}\end{array}\right)\left(\begin{array}{c}
M_{-}\\
M_{+}\end{array}\right)\end{equation}
Thus we get the diagonalized kinetic terms:
\be
\mathcal{L}_{KE}=\frac{1}{2}\tr\left[DM_{+}^{\dagger}DM_{+}\right]+\frac{1}{2}\tr\left[DM_{-}^{\dagger}DM_{-}\right]\;.
\ee
The covariant derivatives for fields $M_{\pm}$ are
\begin{eqnarray}
DM_{\pm} & = & \partial M_{\pm}-ig{W}M_{\pm}+ig'M_{\pm}{B}\;,\\
DM_{\pm}^{\dagger} & = & \partial M_{\pm}^{\dagger}+igM_{\pm}^{\dagger}{W}-ig'{B}M_{\pm}^{\dagger}\;,
\end{eqnarray}
where the contraction with the SU(2) generators is again implicit, for example ${W}=W^a\tau^a$. In the unitary gauge, the spectrum consists of the two $J^P=0^{+}$ particles and three $0^{-}$ pions, while the remaining three pions are absorbed to become the longitudinal components of the gauge bosons by the Higgs mechanism. The pion content of the fields $M_{\pm}$ in the unitary gauge is determined by requiring that the linear coupling between the pions and the gauge bosons vanishes, and that the pions have a correctly normalized
kinetic term. Then
\begin{equation}
M_{\pm}=\frac{1}{\sqrt{2}}\left(s_{\pm}+f_{\pm}\mp2i\frac{f_{\mp}}{v_{w}}\pi\right)\label{eq:Ms} \ ,
\end{equation}
where
\begin{equation}
f_{\pm}\equiv\frac{1}{\sqrt{2}}\left\langle\tr M_{\pm}\right\rangle =\frac{\sqrt{1\pm(c_{3}/\alpha)y_{Q}}}{\sqrt{2}}\left(f\pm v\right)\label{eq:fs} \ .
\end{equation}
and the kinetic part in the unitary gauge is

\begin{eqnarray}
\mathcal{L}_{KE} & = & \frac{1}{2}(\partial s_{-})^{2}+\frac{1}{2}(\partial s_{+})^{2}+\frac{1}{2}(\partial\bar{\pi})^{2}+\frac{1}{8}\overline{V}_Z^2\left[(s_{-}+f_{-})^{2}+(s_{+}+f_{+})^{2}\right]\nonumber \\
 &  & +\frac{1}{2}\overline{V}_A\cdot\bar{\pi}\times\partial\bar{\pi} +\frac{1}{2}\overline{V}_Z\cdot\left[\frac{f_{+}}{v_{w}}\left(\partial s_{-}\bar{\pi}-s_{-}\partial\bar{\pi}\right)-\frac{f_{-}}{v_{w}}\left(\partial s_{+}\bar{\pi}-s_{+}\partial\bar{\pi}\right)\right]\nonumber\\
 &  &\text{+ 4-point interactions}\nonumber
 \label{unitary_KE}
\end{eqnarray}
with 
\begin{equation}
\overline{V}_A \equiv\left(\begin{array}{c}
gW_{1}\\
gW_{2}\\
gW_{3}+g^\prime B\end{array}\right) \ , \qquad  \overline{V}_Z \equiv \left(\begin{array}{c}
gW_{1}\\
gW_{2}\\
gW_{3}-g^\prime B\end{array}\right) \label{def-vectors} \ .
\end{equation}
The constraint to reproduce the $W$ gauge boson mass is:
\begin{equation}
f_{+}^{2}+f_{-}^{2}=v_{w}^{2}\equiv\frac{4m_{w}^{2}}{g^{2}}=f^{2}+v^{2}+2(c_{3}/\alpha)y_{Q}fv\;.
\label{constraint}\end{equation}
The potential is:
\begin{eqnarray}
V_{0} & = & \frac{1}{2}m^{2}_H\left(h^{2}+\bar{\pi}_{H}^{2}\right)+\frac{\lambda_{H}}{4!}\left(h^{2}+\bar{\pi}_{H}^{2}\right)^{2}\nonumber\\
 &  & +\frac{1}{2}m^{2}_M\left(s^{2}+\bar{\pi}_{M}^{2}\right)+\frac{\lambda_{M}}{4!}\left(s^{2}+\bar{\pi}_{M}^{2}\right)^{2}\nonumber\\
 &  & -y_{Q}c_{1}\alpha f^{2}\left(sh+\bar{\pi}_{H}\cdot\bar{\pi}_{M}\right)-y_{Q}\frac{c_{2}\alpha}{6}\left(s^{2}+\bar{\pi}_{M}^{2}\right)\left(sh+\bar{\pi}_{H}\cdot\bar{\pi}_{M}\right)\nonumber\\
 &  & -y_{Q}\frac{c_{4}}{6\alpha}\left(h^{2}+\bar{\pi}_{H}^{2}\right)\left(sh+\bar{\pi}_{H}\cdot\bar{\pi}_{M}\right)
 \label{potential}\end{eqnarray}
In order for the potential (\ref{potential}) to be bounded from below, the couplings $\lambda_{H}$ and $\lambda_{M}$ must be large enough so that the corresponding operators, which are positive definite, dominate over other dimension four operators. In \cite{Ferreira:2004yd}, a necessary but not sufficient condition guaranteeing the tree-level boundness of the  potential was given and reads: 
\be
4y_{Q}|\frac{c_{4}}{\alpha}+c_{2}\alpha|<\lambda_{H}+\lambda_{M}\;.
\label{boundedness}
\ee
We have experimented with this condition and convinced ourselves that it holds for a quite wide choice of parameters in the Lagrangian. Possible violations seem to be confined, for reasonable values of the parameters, within a few percent of the total parameter space. 

The pion mass is calculated by projecting the pion fields in $M$ and $H$ onto the physical pions, and then calculating the mass term of that physical field. The orthogonal, absorbed pion is massless, which will remain true at the one-loop level when we evaluate the effective potential. The potential is diagonalized by writing the $s$ and $h$ fields in terms of a rotation angle, reminiscent of $\tan\beta$ in the MSSM, and physical fields $s_{p}$ and $h_{p}$:
\begin{eqnarray}
\left(\begin{array}{c}
s\\
h\end{array}\right)=\frac{1}{2}\left(\begin{array}{cc}
\frac{1}{\sqrt{1-(c_{3}/\alpha)y_{Q}}} & \frac{1}{\sqrt{1+(c_{3}/\alpha)y_{Q}}}\\
-\frac{1}{\sqrt{1-(c_{3}/\alpha)y_{Q}}} & \frac{1}{\sqrt{1+(c_{3}/\alpha)y_{Q}}}\end{array}\right)\left(\begin{array}{cc}
1 & -1\\
1 & 1\end{array}\right)\left(\begin{array}{cc}
\cos\theta & \sin\theta\\
-\sin\theta & \cos\theta\end{array}\right)\left(\begin{array}{c}
s_{p}\\
h_{p}\end{array}\right)
\end{eqnarray}
The rotation angle is then solved for by requiring the mixing term
to vanish: 

\begin{eqnarray}
\tan2\theta=-\frac{y_{Q}}{\sqrt{\alpha^{2}-y_{Q}^{2}c_{3}^{2}}}\frac{c_{4}v^{2}
+(2c_{1}+c_{2})f^{2}\alpha^{2}+c_{3}(N_{H}+N_{M})}{N_{H}-N_{M}} \ . 
\label{tantheta}
\end{eqnarray}
with 
\[
N_{H}=m_{H}^{2}+\frac{\lambda_{H}v^{2}}{2}-y_{Q}fv\frac{c_{4}}{\alpha}\;,\qquad 
N_{M}=m_{M}^{2}+\frac{\lambda_{M}f^{2}}{2}-y_{Q}fvc_{2}\alpha \ . \] 

\section{Radiative Corrections and Constraints}
\label{analysis}

Now we proceed to calculate radiative corrections to the tree-level model. We begin by describing the evaluation of the one-loop effective potential and then proceed to calculate the FCNC interactions and oblique corrections $S$ and $T$.  After that we describe how the parameter space is scanned. In the final subsection we present the numerical results and constraints. 
\subsection{Coleman-Weinberg effective potential}

The model possesses an elementary and a composite Higgs. The latter is coupled to the SM fermions. In particular the top Yukawa sector leads to large radiative corrections on the effective potential and we want to investigate this effect here. On the other hand the thought composite Higgs sector and its interplay with the elementary one will also be affected by radiative corrections. These should be computed via the full underlying technicolor theory. In the spirit of \cite{Carone:1992rh} we will also consider these corrections. Our approximations for the composite sector are, however, different given that we compute directly the radiative corrections using the effective theory introduced earlier. The analysis of the one-loop effective potential can be also viewed as a test of the model against the radiative corrections. We adopt the 1-loop effective potential derived by Coleman and Weinberg \cite{Coleman:1973jx} which reads:
\begin{equation}
V_{0}+\frac{1}{64\pi^{2}}\left\{ -{\rm Tr}\left[\left(mm^{\dagger}\right)^{2}\ln mm^{\dagger}\right]+3{\rm Tr}\left[M_G^{4}\ln{M_G^2}\right]+{\rm Tr}\left[M^{4}_S\ln M^{2}_S\right]\right\} +\text{infinite terms.}\label{eq:efp}\end{equation}
Here $V_{0}$ is the tree-level potential, $m$ is the fermion mass matrix, $M_G$ is the gauge boson mass matrix and $M_S$ is the scalar boson mass matrix, all written in terms of classical fields with respect to which we will minimize the effective potential. 

To explicitly evaluate the one-loop terms, one must handle two types of infinities. First, there are terms in the one-loop potential that diverge with the cutoff. Second, there are logarithms which will diverge if an eigenvalue approaches zero. The first type of infinities are tamed by renormalization. However, our effective Lagrangian is fully renormalizable in four dimensions, and thus we can simply add counterterms to the tree level Lagrangian. The second type of infinity is cured by choosing suitable renormalization conditions, which amounts to assuming a suitable renormalization scheme. The scheme can be chosen in any convenient way, and we choose the renormalization conditions in such a way that we are able to perform the parameter scan as explained in the subsection \ref{PSM}. We hence choose our renormalization conditions so that  the extrema conditions are not affected by quantum corrections. Our renormalization conditions are as follows: 
\begin{eqnarray}
\left.\frac{\partial V_{1}}{\partial h}\right|_{\left\langle \right\rangle }=0\;,\;\left.\frac{\partial V_{1}}{\partial s}\right|_{\left\langle \right\rangle } &=& 0 \nonumber \\
\left.\frac{\partial^{4}V}{\partial h^{4}}\right|_{\left\langle \right\rangle }=\lambda_{H}\;,\;\left.\frac{\partial^{4}V}{\partial s^{4}}\right|_{\left\langle \right\rangle } &=& \lambda_{M} \nonumber \\ \\
\left.\frac{\partial^{2}V}{\partial h\partial s}-\frac{s^{2}}{2}\frac{\partial^{2}V}{\partial h\partial s^{3}}-\frac{h^{2}}{2}\frac{\partial^{2}V}{\partial h^{3}\partial s}\right|_{\left\langle \right\rangle } &=& -y_{Q}(\alpha c_{1})f^{2}\nonumber \\
\left.\frac{\partial^{4}V}{\partial h\partial s^{3}}\right|_{\left\langle \right\rangle }=-y_{Q}(\alpha c_{2})\;,\;\left.\frac{\partial^{4}V}{\partial h^{3}\partial s}\right|_{\left\langle \right\rangle } &=& -y_{Q}(c_{4}/\alpha)\;,\;\nonumber
\label{eq:rencond}
\end{eqnarray}
 where the renormalization conditions have been evaluated at the extremum and $V_1$ denotes the one-loop contribution.  
 
 We retain the contributions which are relevant, i.e. the effect of the top-quark and the scalars. We neglected, for example, the contributions of the gauge bosons which are suppressed at least by $g^2$.  Thus one must calculate the
eigenvalues of the 8$\times$8 scalar mass matrix. To be consistent with our initial approximations we will not consider terms quadratic in  $y_{Q}$ in the potential.  This approximation reduces the diagonalization to two $4\times 4$ matrices. The conditions above are solved perturbatively in any dimensionless coupling. The one-loop terms may develop an imaginary part in some parts of the parameter space due to the truncation in $y_Q$.

The top-quark Yukawa coupling, which enters the calculation, depends strongly on $v$: $y_{t}=v_w/(\sqrt{2}v)$ at tree level.  In our renormalization scheme the mass squared of the fundamental scalar gets a large  
positive correction from the top of the order of ${4v^{2}y_{t}^{4}}/{(3\pi)}$. This  
is especially important when $v$ is small. Other corrections to the scalar masses are proportional to  
$y_Q$ because of our renormalization conditions that allow  
$m^2_M$ and $m^2_H$ to run and also because we assumed that the  
counterterms were determined perturbatively in the couplings.

\subsection{Flavor Changing Neutral Currents and Oblique Corrections Set Up}

\begin{figure}[h!tb]
\includegraphics[width=15cm]{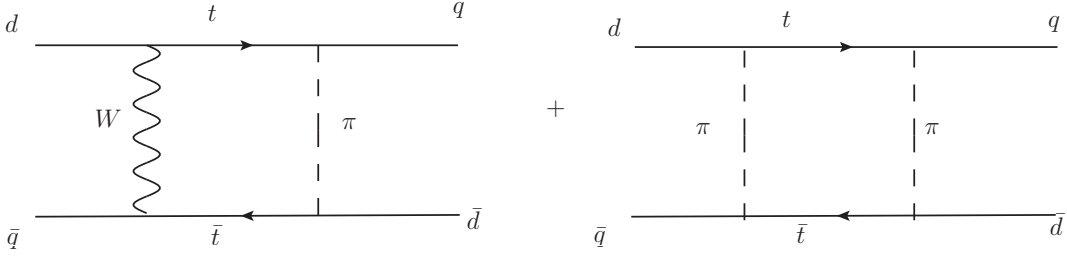}
\caption{Box diagrams contributing beyond the SM to $\Delta q=2$ FCNC interactions for $q=s\,,b$. 
Of the possible quark flavors running in the loops the top quark provides dominant contribution due to its large Yukawa coupling to the scalar degrees of freedom.}
\label{fcnc_diagrams}
\end{figure}

Let us now consider the analysis of the FCNC constraints. We consider $\Delta s,b=2$ mixings for which the physical pions contribute through the box diagrams shown in Fig. \ref{fcnc_diagrams}. These diagrams and corresponding integrals are the ones in \cite{Carone:1992rh}. However, we have included the effects of kinetic mixing in the effective Lagrangian, which complicates the analysis. 

The couplings of the charged physical pions and QCD quarks can be extracted from the Yukawa interactions. Taking into account the CKM mixing matrix, denoted by $V_{\rm{CKM}}$ below, and the kinetic mixing of the scalars, these couplings are
\be
i\left(\frac{f_+/v_w}{2\sqrt{1-(c_{3}/\alpha)y_{Q}}}+\frac{f_-/v_w}{2\sqrt{1+(c_{3}/\alpha)y_{Q}}}\right)\left[\overline{D}_LV_{\rm{CKM}}^\dagger\pi^- h_UU_R+\overline{U}_L\pi^+ V_{\rm{CKM}}h_DD_R+{\textrm{h.c.}}\right] \nonumber\\
\label{couplings}
\ee
Here $U$ ($D$) is a vector of up (down)-type quarks, and $h_U$ ($h_D$) is a diagonal matrix of the corresponding Yukawa couplings. The top loops dominate over up and charm loops due to large top Yukawa coupling. The contributions from the diagrams with one pion and $W$ exchange as well as with two pion exchange are, respectively,
\be
&& g^2\left(\frac{K}{v}\right)^2m_t^4(V_{td}V^\ast_{tb})^2\left[A(m_t,m_\pi)+\frac{1}{4m_W^2}B(m_t,m_\pi)\right],\nonumber \\
&& 2\left(\frac{K}{v}\right)^4m_t^4(V_{td}V^\ast_{tb})^2 C(m_t,m_\pi)
\ee
where $K$ is the prefactor in parentheses in (\ref{couplings}), and the required integrals are \cite{Inami:1980fz}
\be
A(m_1,m_2) &=& \frac{1}{16\pi^2}\left[\frac{m_2^2\ln(m_1^2/m_2^2)}{(m_2^2-m_1^2)^2(m_2^2-M_W^2)}+
\frac{M_W^2\ln(m_1^2/M_W^2)}{(M_W^2-m_1^2)^2(M_W^2-m_2^2)}\right.\\
&&\quad \left.-\frac{1}{(m_1^2-m_2^2)(m_1^2-M_W^2)}\right],\nonumber \\ \nonumber\\
B(m_1,m_2) &=& \frac{1}{16\pi^2}\left[\frac{m_2^4\ln(m_2^2/m_1^2)}{(m_2^2-M_W^2)(m_1^2-m_2^2)^2}+\frac{M_W^4\ln(M_W^2/m_1^2)}{(M_W^2-m_2^2)(m_1^2-M_W^2)^2}\right. \\
&&\quad \left.+\frac{m_1^2}{(m_1^2-m_2^2)(m_1^2-M_W^2)}\right]\nonumber \\\nonumber \\
C(m_1,m_2) &=& \frac{1}{16\pi^2}\left[\frac{m_2^2+m_1^2}{(m_2^2-m_1^2)^2}+\frac{2m_1^2m_2^2}{(m_2^2-m_1^2)^3}\ln\left(\frac{m_1^2}{m_2^2}\right)\right],
\ee

Using the above we can constrain the model through the observed mass difference of neutral $B$ and $K$ mesons and through the usual CP-violation parameter $\epsilon_K$; for the required formulas, see \cite{Fukano:2009zm}. Since the theoretical uncertainties coming from hadronic matrix elements are large  one can see that $\epsilon_K$ provides, de facto, the most stringent bound  and will be used below. We have checked that the two-pion exchange diagram above dominates over the diagram with a single pion exchange practically over the whole parameter space.

Recently, it has been shown \cite{Fukano:2009zm} that one cannot neglect the effects of the heavy vectors, from the technicolor sector, on flavor observables. We are now extending the analysis above to investigate the effect of the heavy vectors. 

Before presenting the actual constraints we will first compute the oblique corrections $S$ and $T$ \cite{Peskin:1990zt}
\be
S &=& -16\pi \Pi_{3Y}^\prime(0),\nonumber \\
T &=& \frac{4\pi}{s_w^2c_w^2 M_Z^2}(\Pi_{11}(0)-\Pi_{33}(0)), 
\ee
 within the present model. To evaluate the corrections we need to estimate the new contributions to the vacuum polarizations
\be
\Pi_{3Y}(q^2),~~\Pi_{ii}(q^2) ~~(i=1,2).
\ee
We use the Feynman rules determined from (\ref{unitary_KE}) and tabulated for completeness in Appendix A.

\begin{figure}[h!tb]
\includegraphics[width=15cm]{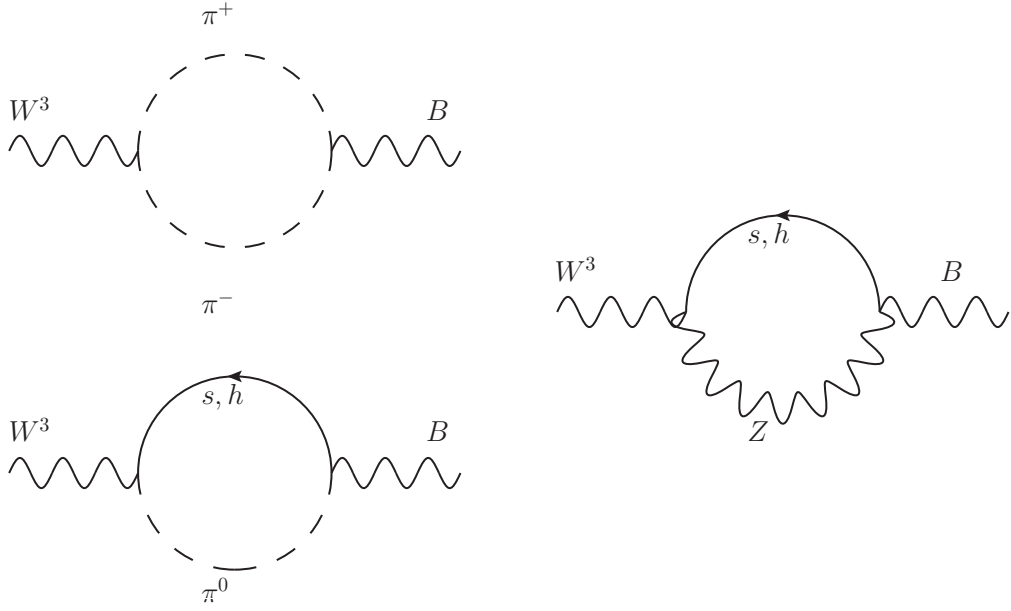}
\caption{The diagrams required for the perturbative evaluation of the vacuum polarization $\Pi_{3Y}$ within the effective theory.}
\label{s_diagrams}
\end{figure}

\begin{figure}[!htb]
\includegraphics[width=15cm]{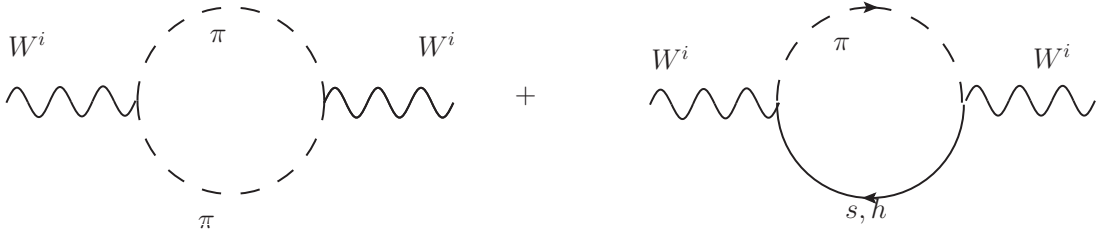}
\caption{Diagrams required for the perturbative evaluation of the vacuum polarization $\Pi_{11}$ within the effective theory. The vacuum polarization $\Pi_{33}$, also needed for $T$-parameter, is obtained from similar diagrams with replacement $W^1\rightarrow W^3$.}
\label{t_diagrams}
\end{figure}

The origin of the $(S,T)$-plane corresponds to the SM with a given value of the mass of the Higgs denoted by $m_{\textrm{ref}}$. We have removed the SM Higgs sector and added new sectors as described in detail in sec. \ref{model}. Then the S-parameter is
\be
S=S_{\textrm{SM}}(m_{\textrm{ref}})-S_H(m_{\textrm{ref}})+S_{\textrm{new}}=S_{\textrm{new}}-S_H(m_{\textrm{ref}}),
\ee
because $S_{\textrm{SM}}(m_{\textrm{ref}})=0$ by definition. Similar considerations apply also to $T$, and using these definitions we obtain finite expressions for $S$ and $T$. We use dimensional regularization in the $\overline{MS}$ scheme. The final result can be expressed in terms of two integrals,
\be
I_1(m_1,m_2,q) &=& \int_0^1 dx\Delta\log\frac{\mu^2}{\Delta},\nonumber \\
I_2(m_1,m_2,q) &=& \int_0^1 dx m_1^2\log\frac{\mu^2}{\Delta},
\ee
where $\Delta\equiv\Delta(m_1,m_2,q)=xm_2^2+(1-x)m_1^2-x(1-x)q^2$ and $\mu$ is an arbitrary mass scale. Furthermore, for notational convenience we define
\be
C_h &=& \frac{1}{\sqrt{2}v_w}(\cos\theta(f_+-f_-)+\sin\theta(f_-+f_+)),\nonumber \\
C_s &=& \frac{1}{\sqrt{2}v_w}(\cos\theta(f_-+f_+)+\sin\theta(f_--f_+)),
\label{oblique_constants}
\ee
so that $C_h^2+C_s^2=1$ and $f_{\pm}$  and the mixing angles are defined in Sec. \ref{model}, see Eqs. (\ref{eq:fs}) and (\ref{tantheta}). The diagrams required for the evaluation of the vacuum polarization $\Pi_{3Y}$ contributing to the $S$-parameter are shown in Fig. \ref{s_diagrams}, and the diagrams contributing to the vacuum polarizations $\Pi_{11}$ and $\Pi_{33}$, affecting the $T$-parameter, are shown in Fig. \ref{t_diagrams}. 


With these preliminary definitions we have
\be
\Pi_{3Y}(q^2) &=& \frac{1}{32\pi^2}\left[-C_h^2\left(I_1(m_\pi,m_s)-2I_2(M_Z,m_h)+I_1(M_z,m_h)\right)\right.\nonumber \\
& & -C_s^2\left(I_1(m_\pi,m_h)-2I_2(M_Z,m_s)+I_1(M_Z,m_s)\right)\\
& & \left.+I_1(m_\pi,m_\pi)+I_1(M_Z,m_{\textrm{ref}})-2I_2(M_Z,m_{\textrm{ref}})\right],\nonumber
\ee
where we have dropped all $q$-independent contributions since we need only the $q^2$-derivative of this quantity. Note that here and later $m_s$ and $m_h$ are the masses of the physical scalars $s_p$ and $h_p$. The contributions to the correlators needed for the $T$-parameter are obtained similarly, and the result is
\be
\Pi_{11}(q^2) &=& \frac{1}{32\pi^2}\left[C_h^2\left(I_1(M_W,m_h)+I_1(m_\pi,m_s)-2I_2(M_W,m_h)\right)\right.\nonumber \\
& & +C_s^2\left(I_1(M_W,m_s)+I_1(m_\pi,m_h)-2I_2(M_W,m_s)\right)\nonumber \\
& & +I_1(m_\pi,m_\pi)-I_1(M_W,m_{\textrm{ref}})+2I_2(M_W,m_{\textrm{ref}})\\
& & \left.+\frac{3}{2}m_\pi^2-\frac{1}{2}m_{\textrm{ref}}^2+\frac{1}{2}m_s^2+\frac{1}{2}m_h^2-\frac{1}{3}q^2\right],\nonumber
\ee
and
\be
\Pi_{33}(q^2)=\Pi_{11}(q^2) \quad \text{with}\quad M_W\rightarrow M_Z.
\ee
We now have all the ingredients to constrain the model. 

\subsection{Parameter Scan Method}
\label{PSM}
For a scan of the parameter space, we use the following procedure: first, we generate thousands of values of the couplings $\lambda_{H}$, $\lambda_{M}$, and $y_{Q}$ within the following parameter range  $(0.01 - 150)$ for the first two parameters and $(0.01 - 12)$ for $y_Q$; second, we allow each of the parameters $c_{2}$... $c_{4}$ to take the two values +1 or -1; third, we allow for $\alpha$ the value of $\pi$ or $4\pi$. Finally, we generate thousands of values for $v$ and solve for $f$ using  (\ref{constraint}). The range of $v$ is defined in such a way that $f$ is always a real quantity and $y_t<4\pi$.  At this point we use the extrema conditions to read off the parameters $m^{2}_H$ and $m^{2}_M$. Extremizing the potential (\ref{potential}) gives us
\be
0 &=& m^{2}_Hv+\frac{v^{3}\lambda_{H}}{6}-y_{Q}\frac{c_{4}}{2\alpha}fv^{2}-y_{Q}\alpha c_{1}f^{3}-y_{Q}\frac{\alpha c_{2}}{6}f^{3}
\label{firstextremum} \\
0 &=& m^{2}_Mf+\frac{f^{3}\lambda_{M}}{6}-y_{Q}\frac{c_{4}}{6\alpha}v^{3}-y_{Q}\alpha c_{1}f^{2}v-y_{Q}\frac{\alpha c_{2}}{2}f^{2}v.
\label{extremum}
\ee
Now we have a full set of parameters, and are in the position to determine the pion mass $m^{2}(\pi)$ as well as the masses of the scalars $m^{2}_h$, $m^{2}_s$ and the associated mixing angle $\theta$ in terms of the other model parameters. In order to be at a local minimum, we checked that all physical masses are positive.

However, in addition to the minima we find in the way depicted above, there might be other minima with a lower value of the potential \cite{Barroso:2007rr} and a better algorithm would be advisable which is beyond the scope of this work. 

Finally, the parameter space of the model deserves a remark. The equation
(\ref{constraint}) is a quadratic equation for $f$, with the solutions
\[
f(v)\rightarrow-vy_{Q}\frac{c_{3}}{\alpha}\pm\sqrt{v_{w}^2-v^2+\left(vy_{Q}\frac{c_{3}}{\alpha}\right)^{2}}\]

Under a $Z_{2}$ symmetry of the model, $M\rightarrow-M$, $H\rightarrow-H$, the two roots are exchanged and thus we can span the whole parameter space with only the positive root. Furthermore, because the model is also symmetric under $c_i\rightarrow-c_i$ and $H\rightarrow-H$, we can span the whole parameter space while limiting ourselves to $v^2<v_w^2/(1-(c_{3}y_Q/\alpha)^2)$, the positive root for $f(v)$ and $c_1>0$.

\subsection{Numerical results: constraining the model}

We will now constrain the parameter space by requiring compatibility with oblique measurements, direct searches of the Higgs boson, and FCNCs. We will see that these constraints provide stringent cuts on the parameter space.

For the oblique corrections, we demand that $S$ and $T$ do not deviate substantially from the reported fits. We note that there exists two extensive fits performed by the LEP Electroweak Working Group (LEPEWWG) \cite{:2005ema} and independently by the PDG \cite{Amsler:2008zzb}. Both fits find that the SM, defined to lie at $(S,T)=(0,0)$ with $m_t=170.9$ GeV and $m_H=117$ GeV, is within $1\sigma$ of the central value of the fit. The two fits disagree on the central best-fit value: LEPEWWG finds a central value $(S,T)=(0.04,0.08)$ while including the low energy data the PDG  finds $(S,T)=(-0.04,0.02)$.  Since the actual level of coincidence inferred from these fits depends on the precise nature of the fit, we allow a broader range of $S$ and $T$ values, roughly corresponding to the $3\sigma$ contour. 

To evaluate the intrinsic $S$ of the composite sector, we use the naive estimate of one loop of techniquarks possessing a large mass with respect to the weak gauge boson one. For the underlying theory with three technicolors and two flavors in the sextet representation this naive estimate gives $S_{\textrm{naive}}=1/\pi$. Since the underlying technicolor sector can be near conformal, its contribution to the true value $S$ is expected to be smaller \cite{Appelquist:1998xf,Sundrum:1991rf}, roughly $\sim 0.7S_{\textrm{naive}}$.  The full $S$ to be compared to experiments contains also the contribution  from the other scalar sector as well as the interplay among the two. 

The intrinsic technicolor contribution to $S$ could also be evaluated via the low energy technihadron spectrum. The latter is linked to the UV properties of the underlying gauge theory via dispersion relations. If one assumes, for example, resonance saturation of the relevant two-point function the bulk of the technicolor $S$ value would then correspond to the one obtained via the composite scalar (if light), vector and axial (possibly other higher spin) resonances exchange. 

We consider two  limits when comparing our model with the experimental value of $S$. First, if the heavy vectors are completely decoupled, one can compute the $S$ contribution from the composite Higgs within the effective theory. This special limit of the parameter space corresponds to the generic two Higgs doublet model. Second if the vectors do not decouple their effect is estimated by adding $S_{\textrm{naive}}=1/\pi$ on the top of the one computed within the effective theory. In this case we are overestimating the contribution from the technicolor sector. 

Our perturbative treatment of the model has some intrinsic uncertainty due to the chiral expansion which we have used in writing down the initial Lagrangian. We have truncated the expansion to include no higher than dimension four operators. However, at dimension six there would appear contributions to both $S$ and $T$ through operators
\be
\frac{c_5}{2\Lambda^2}{\textrm{Tr}}(M^\dagger W^{\mu\nu}MB_{\mu\nu}),
\ee
and
\be
\frac{c_6}{4\Lambda^2}|{\textrm{Tr}}(HY_Q D^\mu M^\dagger)|^2
\ee
respectively. However, due to their suppression by $\alpha^2$ we take their contribution to be negligible in our analysis. Since we have considered only the leading contributions in the Yukawa couplings $y_Q$, we have not taken into account the possible splitting in the masses of the neutral and charged pions which would arise from the operator
\be
\frac{c_7}{2}{\textrm{Tr}}(HY_Q M^\dagger HY_Q M^\dagger)
\ee
which is of dimesion four but also of ${\mathcal{O}}(Y_Q^2)$ and hence neglected in our analysis. 

\begin{figure}[h!tb]
\hspace{-0.5cm}
\includegraphics[width=.45\textwidth]{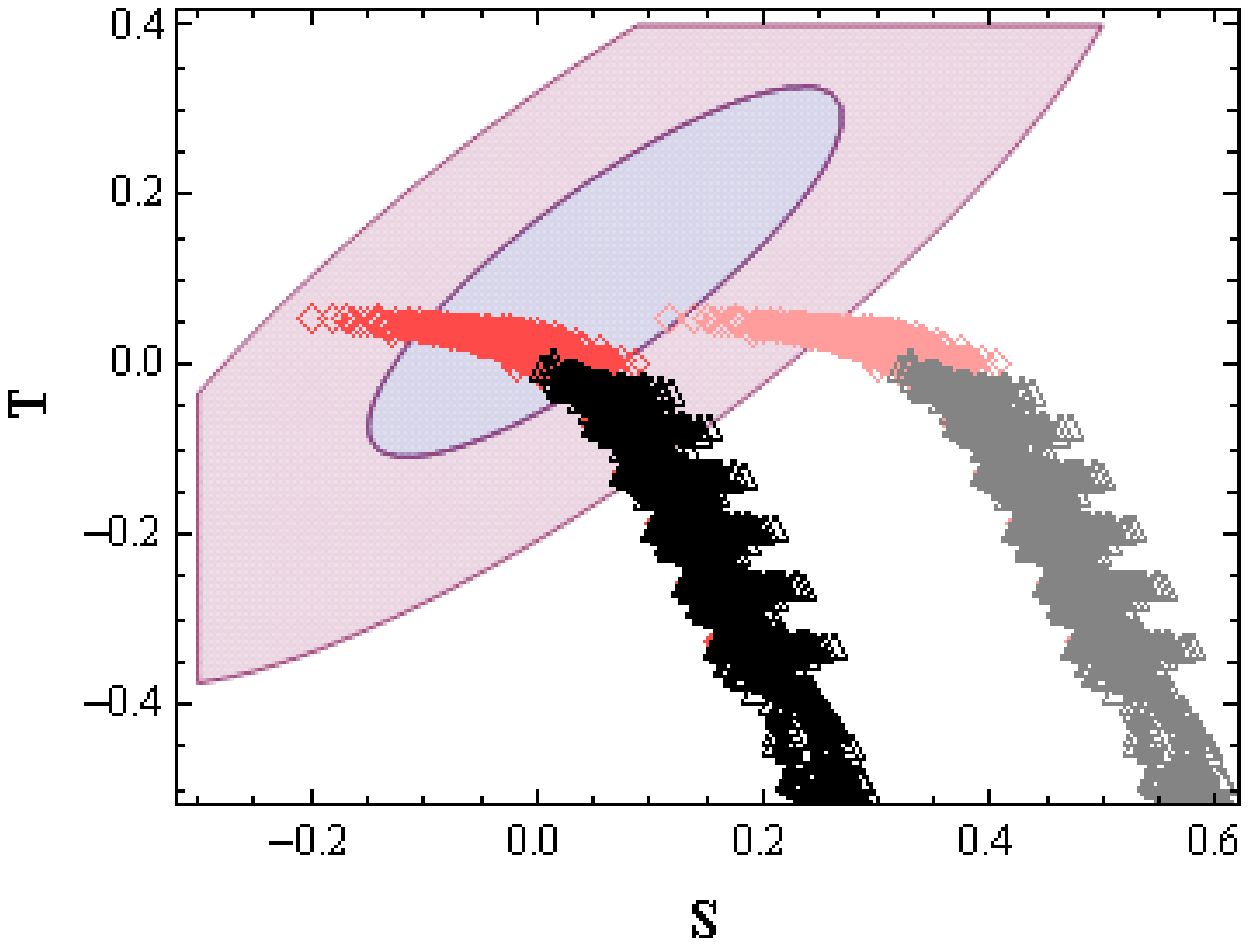}
%\end{minipage}
\hspace{0.5cm}
\includegraphics[width=.55\textwidth]{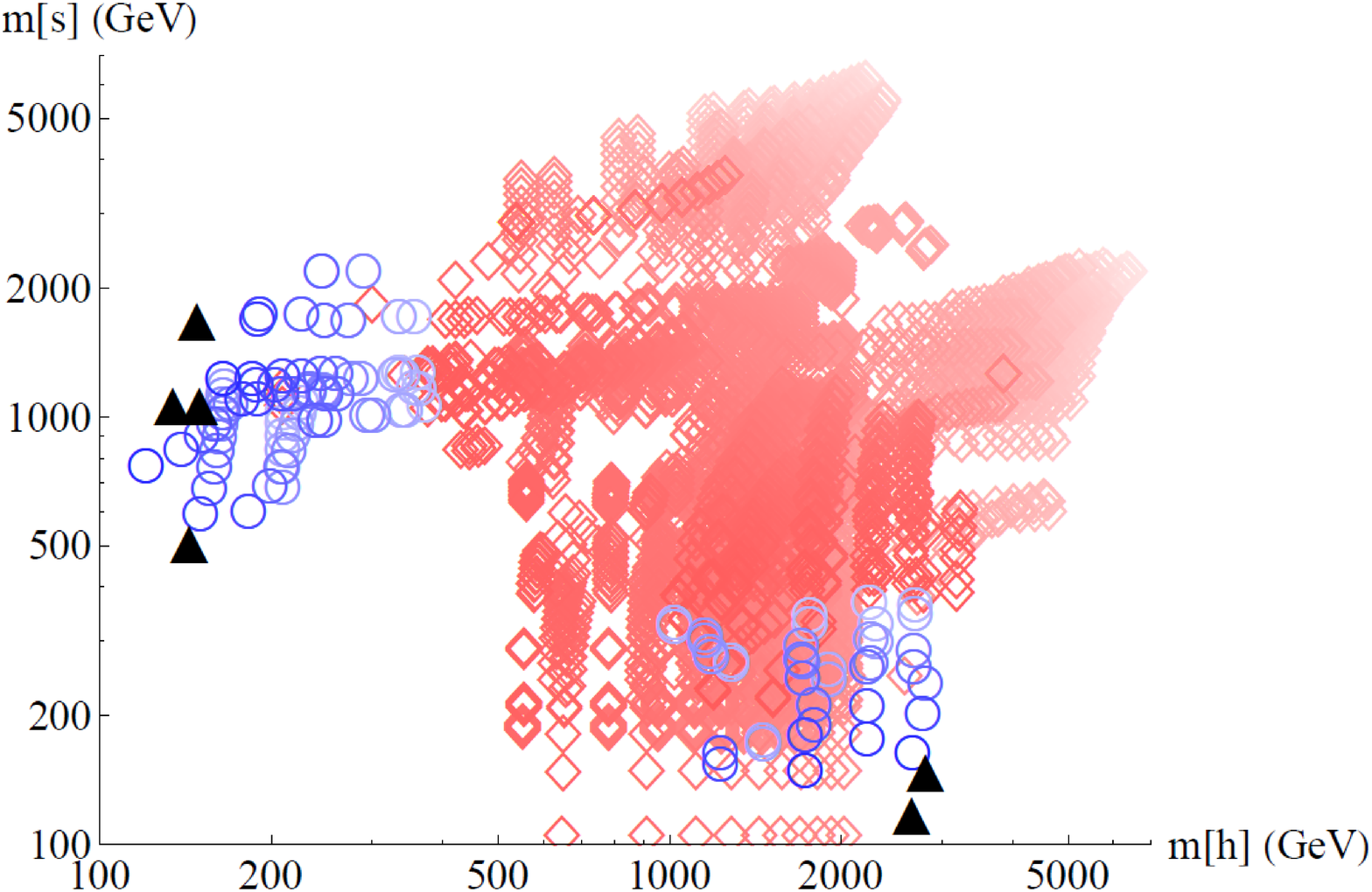}
%\end{minipage}
\caption{Left: The results of the model and the 90\% confidence limit contour allowed by all electroweak data for $m_{\rm ref}=115$ GeV. The light red diamonds are excluded by direct observatons while the black triangles are not. Right: Black triangles show the points consistent with the 90\% S-T confidence limit, blue circles correspond to triangles in the left panel that are within the larger ellipse and the red diamonds to triangles even farther out. Lighter points are also farther out.}
\label{ST_plot}
\end{figure}

The left panel of Fig. \ref{ST_plot} is an S-T plot with the origin corresponding to the SM with a Higgs mass $m_{\textrm{ref}}=117$ GeV. All points pass the FCNC requirements, but the light red diamonds correspond to parameter values that are excluded on the basis of direct searches, soon to be discussed. The leftmost point are the result of a calculation within the effective theory. To these we add $S_{\rm{naive}}$, representing the effects of vectors, to obtain the rightmost set of points. In the forthcoming analysis we concentrate on the leftmost set of points.

The right panel of Fig. \ref{ST_plot} shows the projection of the data set presented in the left panel onto the ($m_h,m_s)$ plane. In this figure the black triangles denote the points consistent with the 90\% confidence contour of the S-T plot. The blue circles correspond to  the points within the larger ellipsis of the S-T plot. Finally, the light red diamonds correspond to points still farther out from the ellipsis.  

It is possible that the lighter scalar could evade detection even below the LEP direct search limit $m_H>114$ GeV \cite{Amsler:2008zzb}, if its coupling to two Z-bosons is suppressed. Here restrictions on oblique corrections and direct search limits intertwine, because we find that for those points with the most compatible S and T values, the coupling is not suppressed. This is seen by looking at both panels of Fig. \ref{ST_plot}. In the left panel, the lighter points are those that are removed by the LEP direct search limits. These disallowed points are the most promising area in terms of S and T values.  On the right panel, we see that the points with the best S and T values are those with at least one light scalar, but for which the light scalar must also satisfy $m>115$ GeV, thus indicating that there is no suppression of the scalar-ZZ vertex for these points. 

The constraint coming from the lightest scalar decay into  $b\bar{b}$ has been taken into account in the simple approximation in which the coupling is the same as in the SM.  This leads to an overestimate of the lightest scalar decay width into fermions. This means that the allowed regions would extend to lower values of the physical scalar masses, especially for the mostly composite scalar, $s$. Although we have here considered $m_{\textrm{ref}}=117$ GeV, the result in the right panel of Fig. \ref{ST_plot} is practically independent of this reference mass value. 

\begin{figure}[h!tb]
\hspace{-0.5cm}\includegraphics[width=8cm]{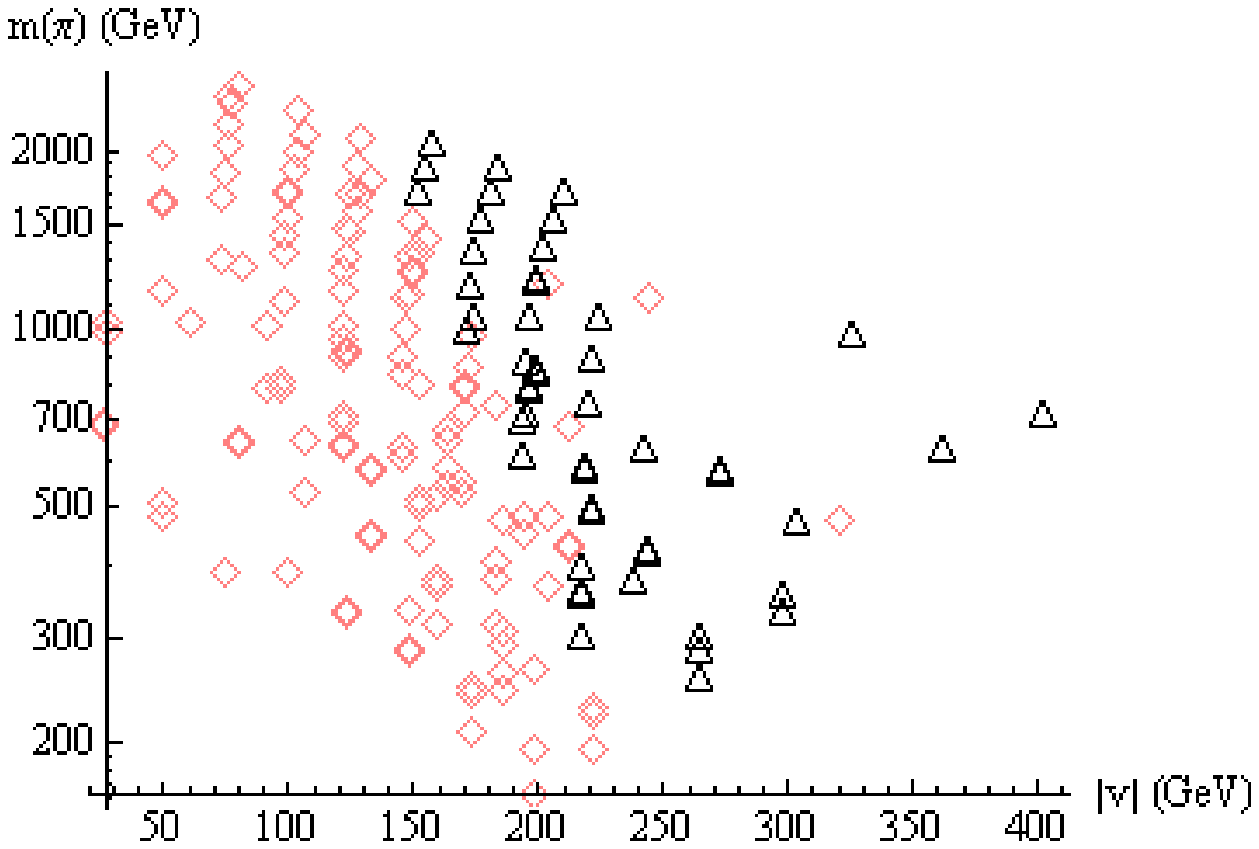}
\includegraphics[width=8cm]{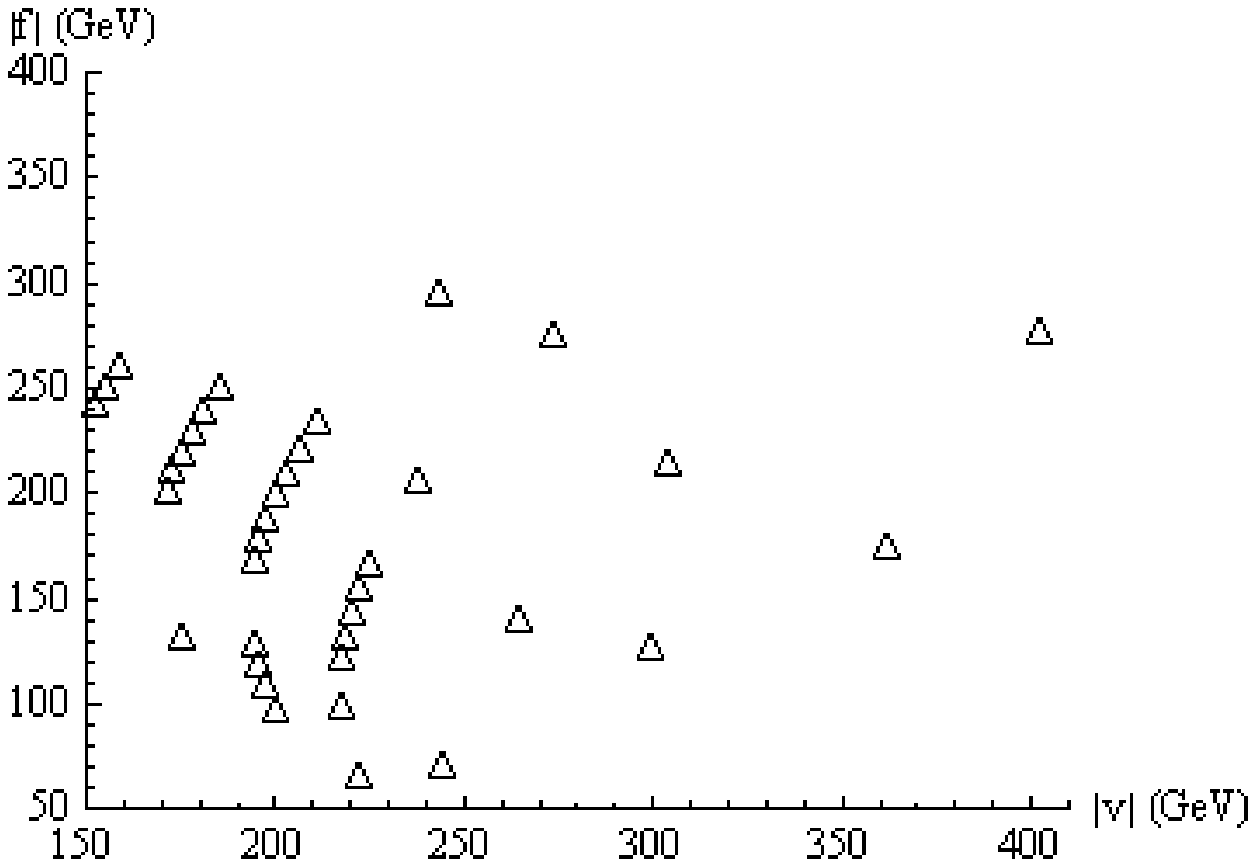}
\caption{Left: The FCNC constraints on parameters $m_\pi$ and $v$ on points satisfying direct search and S-T 90\% confidence limit. Light red diamonds are unallowed, while black triangles are allowed. Right: The allowed values of the condensates $f$ and $v$ after taking all constraints into account.}
\label{mpi_constraint}
\end{figure}

Finally, imposing the FCNC constraints forbids the area of parameter space where the value of the condensate $v$ and the mass of the physical pion $m_{\pi}$ are both small. This is evident when looking at the left panel of Fig. \ref{mpi_constraint}. In the right panel of Fig. \ref{mpi_constraint} we show the allowed values of the condensates $f$ and $v$.  We see that the allowed region for these parameters for each condensate changes roughly in the range of $50$ to $400$ GeV.  

\section{Discussion and outlook}
\label{checkout}

In this paper we have investigated a simple framework to study the interplay between two sources of electroweak symmetry breaking corresponding to a composite and a fundamental Higgs sector.  The presence of the new scalar allows for the generation of mass of any fermion without necessarily invoking new strong dynamics, a la ETC, for which satisfactory models are hard to construct \cite{Appelquist:2003hn}. Of course, there is a price for this, i.e. the Yukawa couplings  remain devoid of a more fundamental explanation.

We constrained the parameter space of the model using the LEP direct search limits, FCNC results and the electroweak precision tests. We find that the model is viable and can be seen as the stepping stone for a well defined extension of the SM featuring a complete solution to both the origin of spontaneous breaking the electroweak symmetry and the mass of any other SM fermion. The model can be seen also as an effective low energy description of a more natural model aimed at explaining the mass of the ordinary fermions. At this point one can embark in a serious study of possible signatures at colliders as done, for example, in \cite{Belyaev:2008yj}. 

\acknowledgments
We thank C. Carone, S. Di Chiara, M.T. Frandsen and E. Simmons for discussions and comments on the manuscript. 

\appendix
\section{Unitary Gauge Feynman rules}

In this appendix we tabulate the basic vertices 
%from which one can derive the Feynman rules 
needed for the processes investigated in this work.

\begin{minipage}[c]{0.2\textwidth}
\includegraphics[width=1.0\textwidth]{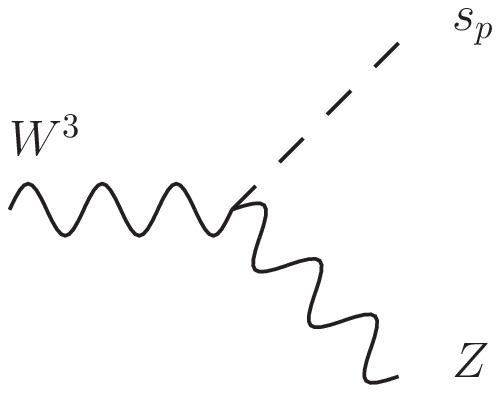}
\end{minipage}
\begin{minipage}[c]{0.2\textwidth}
\[ =i\frac{v_wg^2C_s}{2\cos\theta_w}, \]
\end{minipage}
\begin{minipage}[c]{0.2\textwidth}
\includegraphics[width=1.0\textwidth]{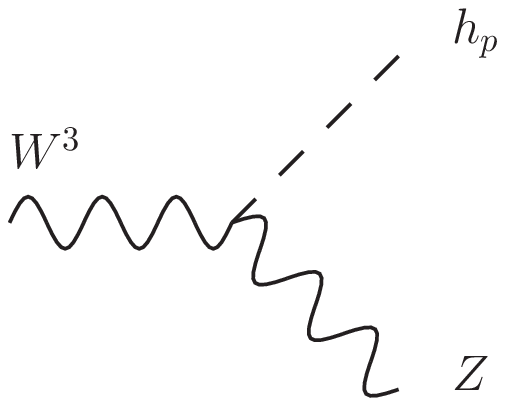}
\end{minipage}
\begin{minipage}[c]{0.2\textwidth}
\[ =i\frac{v_wg^2C_h}{2\cos\theta_w}, \]
\end{minipage}\\
\begin{minipage}[c]{0.2\textwidth}
\includegraphics[width=1.0\textwidth]{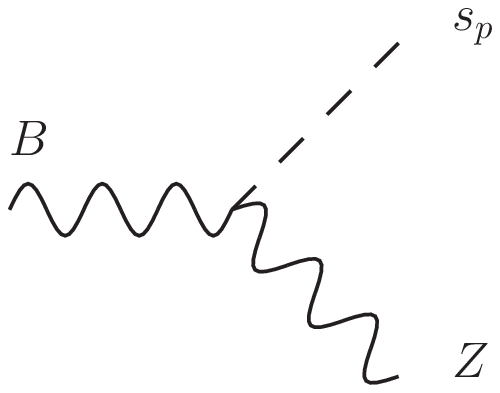}
\end{minipage}
\begin{minipage}[c]{0.2\textwidth}
\[ =-i\frac{v_wg'^2C_s}{2\sin\theta_w}, \]
\end{minipage}
\begin{minipage}[c]{0.2\textwidth}
\includegraphics[width=1.0\textwidth]{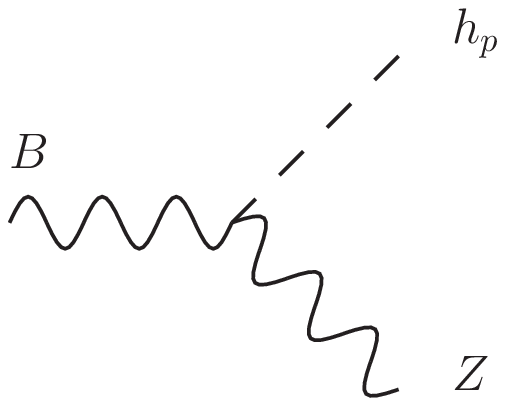}
\end{minipage}
\begin{minipage}[c]{0.2\textwidth}
\[ =-i\frac{v_wg'^2C_h}{2\sin\theta_w},\]
\end{minipage}\\
\begin{minipage}[c]{0.2\textwidth}
\includegraphics[width=1.0\textwidth]{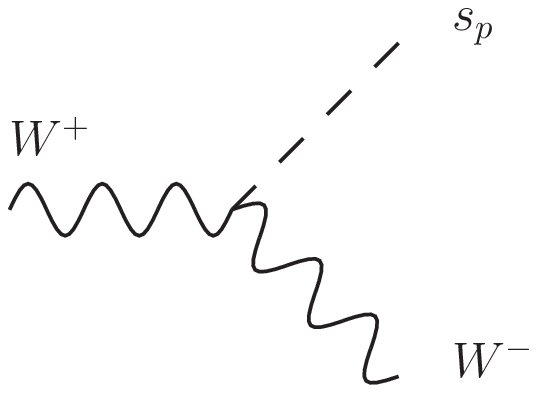}
\end{minipage}
\begin{minipage}[c]{0.2\textwidth}
\[ =i\frac{v_wg^2C_s}{2}, \]
\end{minipage}
\begin{minipage}[c]{0.2\textwidth}
\includegraphics[width=1.0\textwidth]{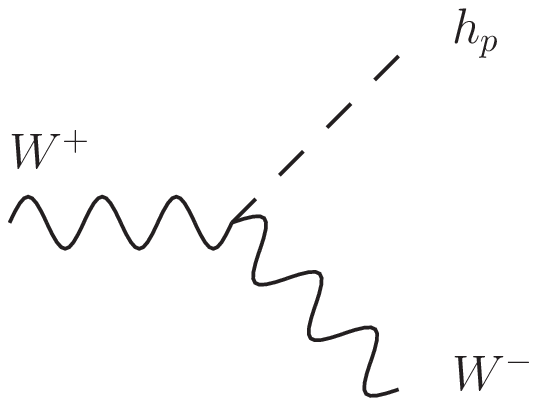}
\end{minipage}
\begin{minipage}[c]{0.2\textwidth}
\[ =i\frac{v_wg^2C_h}{2},\]
\end{minipage}\\
\begin{minipage}[c]{0.2\textwidth}
\includegraphics[width=1.0\textwidth]{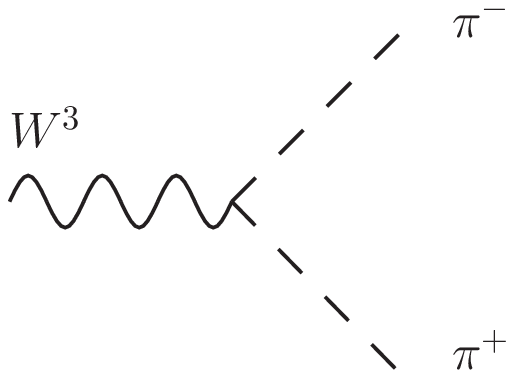}
\end{minipage}
\begin{minipage}[c]{0.2\textwidth}
\[ =i\frac{g}{2}(k_\mu^--k_\mu^+), \]
\end{minipage}
\begin{minipage}[c]{0.2\textwidth}
\includegraphics[width=1.0\textwidth]{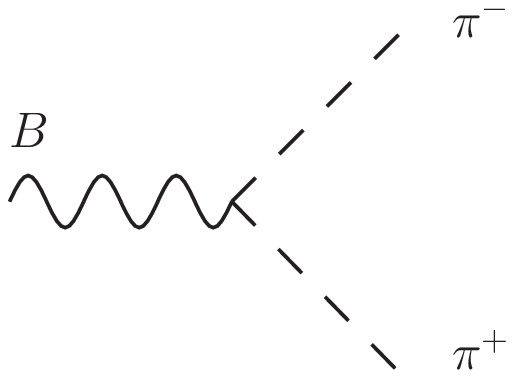}
\end{minipage}
\begin{minipage}[c]{0.2\textwidth}
\[ =i\frac{g'}{2}(k_\mu^--k_\mu^+),\]
\end{minipage}\\
\begin{minipage}[c]{0.2\textwidth}
\includegraphics[width=1.0\textwidth]{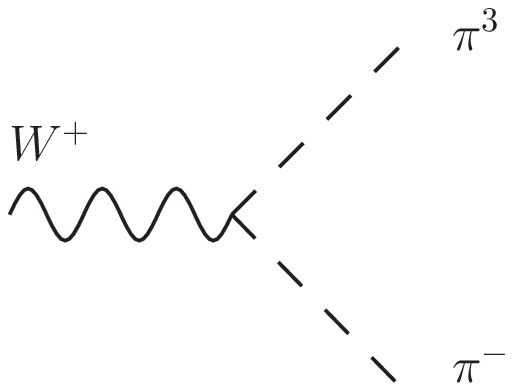}
\end{minipage}
\begin{minipage}[c]{0.2\textwidth}
\[ =i\frac{g}{2}(k_\mu^3-k_\mu^-), \]
\end{minipage}
\begin{minipage}[c]{0.2\textwidth}
\includegraphics[width=1.0\textwidth]{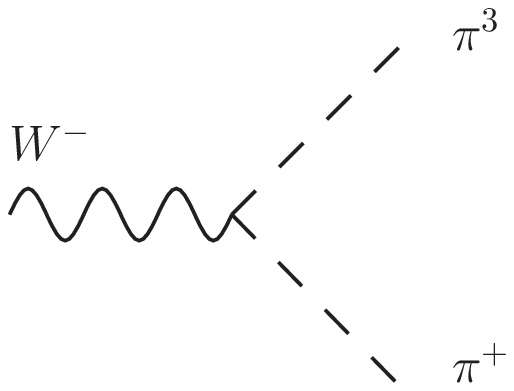}
\end{minipage}
\begin{minipage}[c]{0.2\textwidth}
\[ =i\frac{g}{2}(k_\mu^+-k_\mu^3),\]
\end{minipage}\\
\begin{minipage}[c]{0.2\textwidth}
\includegraphics[width=1.0\textwidth]{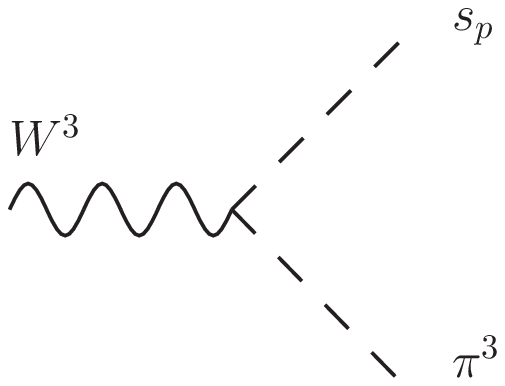}
\end{minipage}
\begin{minipage}[c]{0.2\textwidth}
\[ =i\frac{gC_h}{2}(k_\mu^s-k_\mu^3), \]
\end{minipage}
\begin{minipage}[c]{0.2\textwidth}
\includegraphics[width=1.0\textwidth]{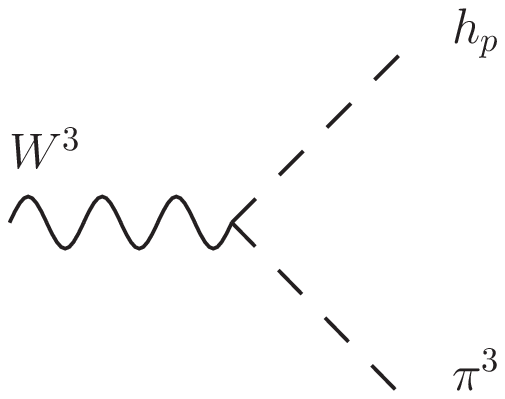}
\end{minipage}
\begin{minipage}[c]{0.2\textwidth}
\[ =i\frac{gC_s}{2}(k_\mu^3-k_\mu^h),\]
\end{minipage}\\
\begin{minipage}[c]{0.2\textwidth}
\includegraphics[width=1.0\textwidth]{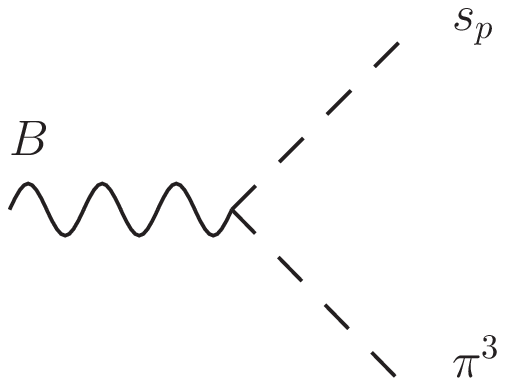}
\end{minipage}
\begin{minipage}[c]{0.2\textwidth}
\[ =i\frac{g'C_h}{2}(k_\mu^3-k_\mu^s), \]
\end{minipage}
\begin{minipage}[c]{0.2\textwidth}
\includegraphics[width=1.0\textwidth]{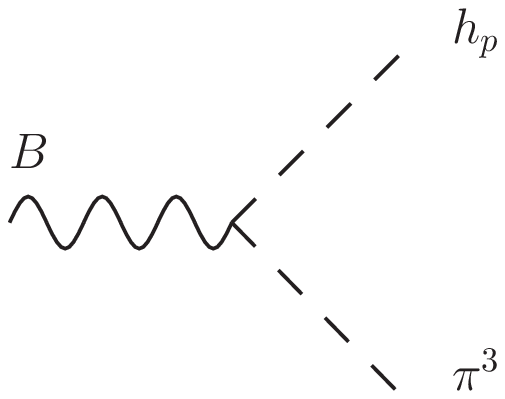}
\end{minipage}
\begin{minipage}[c]{0.2\textwidth}
\[ =i\frac{g'C_s}{2}(k_\mu^h-k_\mu^3),\]
\end{minipage}\\
\begin{minipage}[c]{0.2\textwidth}
\includegraphics[width=1.0\textwidth]{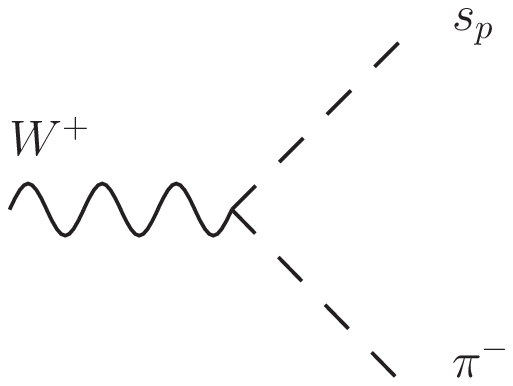}
\end{minipage}
\begin{minipage}[c]{0.2\textwidth}
\[ =i\frac{gC_h}{2}(k_\mu^--k_\mu^s), \]
\end{minipage}
\begin{minipage}[c]{0.2\textwidth}
\includegraphics[width=1.0\textwidth]{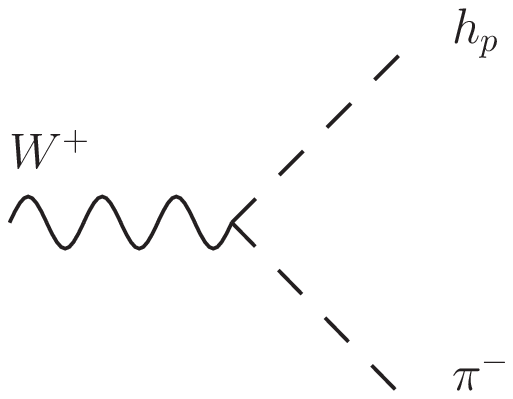}
\end{minipage}
\begin{minipage}[c]{0.2\textwidth}
\[ =i\frac{gC_s}{2}(k_\mu^h-k_\mu^-),\]
\end{minipage}\\
\begin{minipage}[c]{0.2\textwidth}
\includegraphics[width=1.0\textwidth]{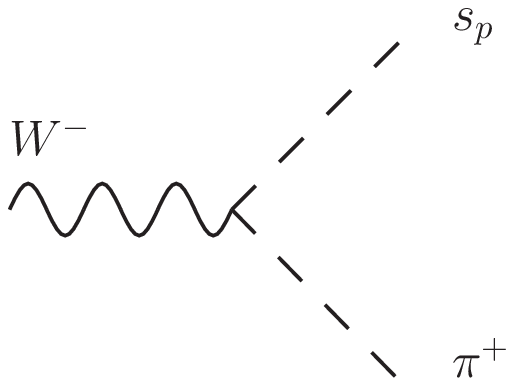}
\end{minipage}
\begin{minipage}[c]{0.2\textwidth}
\[ =i\frac{gC_h}{2}(k_\mu^+-k_\mu^s), \]
\end{minipage}
\begin{minipage}[c]{0.2\textwidth}
\includegraphics[width=1.0\textwidth]{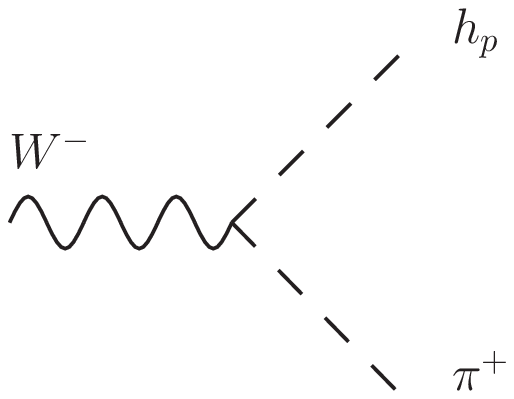}
\end{minipage}
\begin{minipage}[c]{0.2\textwidth}
\[ =i\frac{gC_s}{2}(k_\mu^h-k_\mu^+),\]
\end{minipage}\\

\end{document}